\documentclass[floatfix,aps,prl,reprint]{revtex4-2}
\usepackage{graphicx}
\usepackage{dcolumn}
\usepackage{bm}
\usepackage{braket}
\usepackage{tikz}
\usetikzlibrary{decorations.markings}
\usetikzlibrary{decorations.pathreplacing, arrows.meta}
\usetikzlibrary{positioning}
\usepackage{amsmath}
\usepackage{amsfonts}
\usepackage{hyperref}
\usepackage{pgfplots}
\usepackage{filecontents}

\definecolor{Red}{RGB}{255,0,0}
\definecolor{Green}{RGB}{16,130,46}
\definecolor{Blue}{RGB}{23, 113, 191}

\hypersetup{
	colorlinks,
	linkcolor={red!50!black},
	citecolor={blue!80!black},
	urlcolor={blue!50!black}
}

\newcommand{\pder}[2][]{\frac{\partial#1}{\partial#2}}

\newcommand{\measV}[1]{\mathrm{d}^{3} {#1}}
\newcommand{\meas}[1]{\mathrm{d}{#1}}

\usetikzlibrary{arrows,decorations.pathmorphing,hobby,patterns}
\tikzset{>=latex}
\tikzset{atom/.pic={
\fill[Red](0,0) circle (0.1);
	\foreach \i in {0,45,90,135}
		{	
			\draw[Blue,rotate = \i] (0,0) ellipse (0.5 and 0.25);
		}
	}
}

\tikzset{blob/.pic={
\coordinate (v1) at (-1.5,-1);
\coordinate (v2) at (-1.5,0);
\coordinate (v3) at (-1.5,0.5);
\coordinate (v4) at (-0.7,0.5);
\coordinate (v5) at (-0.5,0);
\coordinate (v6) at (-1,-1);

\fill[draw,use Hobby shortcut,closed=true,fill = Blue!20]
(v1) .. (v2) ..(v3) ..(v4) ..(v5) .. (v6) ;
}
}

\tikzset{arrowLeft/.pic={
\draw[Red] (0,0) -- (1,1) -- (1,0.5) -- (2,0.5) -- (2,-0.5) -- (1,-0.5) -- (1,-1) --cycle ;
}}

\tikzset{arrowRight/.pic={
\draw[Blue] (0,0) -- (-1,1) -- (-1,0.5) -- (-2,0.5) -- (-2,-0.5) -- (-1,-0.5) -- (-1,-1) --cycle ;
}}

\tikzset{funnel/.pic={
\def\diskH{0.6}
\def\R{2}
\draw[fill = Blue!20] (\diskH,0) ellipse (1 and 2);

\fill[fill = Blue!20] (0,-\R) rectangle (\diskH,\R);
\draw[fill = white] (0,-\R) -- (\diskH,-\R);
\draw[fill = white] (0,\R) -- (\diskH,\R);
\draw[fill = Blue!50] (0,0) ellipse ({0.5*\R} and \R);
\fill[fill = Blue!20] (0,0) ellipse ({0.5*0.75*\R} and {0.5*1.7*\R});
\draw[line width = 0.1] (\diskH,{1.1*0.5*0.6*\R}) -- (0,0.8*\R);
\draw[line width = 0.1] (\diskH,{-1.1*0.5*0.6*\R}) -- (0,-0.8*\R);
\draw[line width = 0.1] (0.4*\diskH,0) -- (-0.9*0.5*0.75*\R,0);

\clip (0,0) ellipse ({0.5*0.75*\R} and {0.5*1.7*\R});
\draw[fill = white](\diskH,0) ellipse ({0.5*0.3*\R} and {0.5*0.6*\R});

}}

\tikzset{cylinder/.pic={
\def\diskH{0.6}
\def\R{2}
\draw[fill = Blue!10] (\diskH,0) ellipse (1 and 2);

\fill[fill = Blue!10] (0,-\R) rectangle (\diskH,\R);
\draw[fill = white] (0,-\R) -- (\diskH,-\R);
\draw[fill = white] (0,\R) -- (\diskH,\R);
\draw[fill = Blue!30] (0,0) ellipse ({0.5*\R} and \R);

}}

\begin{document}


\title{Algorithmic Discovery of Casimir-Polder Forces: Repulsion in the Ground State}

\author{Romuald Kilianski}
\email{r.kilianski.1@research.gla.ac.uk}
\author{Claire M. Cisowski}
\author{Robert Bennett}%
\affiliation{School of Physics and Astronomy, University of Glasgow, Glasgow, G12 8QQ UK}

\date{\today}

\begin{abstract}
We present a general-purpose algorithm for automatic production of a structure that induces a desired Casimir-Polder force. As a demonstration of the capability and wide applicability of the method, we use it to develop a geometry that leads to a repulsive Casimir-Polder force on a ground-state atom. The results turn out to be reminiscent of the ring-like geometries previously used to induce repulsion, but with some new features and -- importantly -- discovered completely independently of any input from the user. This represents a powerful new paradigm in the study of atom-surface forces --- instead of the user testing various geometries against a desired figure of merit, the goal can be specified and then an appropriate geometry created automatically. 
\end{abstract}

\maketitle

One of the most remarkable predictions of quantum electrodynamics is the existence of the fluctuating ground state of the electromagnetic field (see, e.g., \cite{milonni2013quantum}). Macroscopic objects impose boundary conditions on this field leading to observable phenomena, the most famous of which is the celebrated Casimir effect \cite{casimirAttractionTwoPerfectly1948}. In its original guise, the Casimir effect was a theoretical (and even philosophical) curiosity, with increasingly elaborate observations guiding an entire sub-field of precision force metrology \cite{sparnaayAttractiveForcesFlat1957,lamoreauxDemonstrationCasimirForce1997,mohideenPrecisionMeasurementCasimir1998,deccaTestsNewPhysics2007,mundayMeasuredLongrangeRepulsive2009,garrettMeasurementCasimirForce2018}. Through the pioneering work of Lifshitz and collaborators \cite{Lifshitz1956,Lifshitz}, theoretical understanding of Casimir force was extended to encompass complex media, geometries and temperature-dependence. In a modern language, the Casimir force is simply one of a collection of phenomena known as dispersion forces \cite{buhmann2012Book1}.

A closely-related effect is the Casimir-Polder (CP) force, whereby a neutral atom or molecule interacts with a polarizable medium via fluctuations of the electromagnetic field \footnote{It is crucial for us to point out that we refer to any interaction between a polarizable atom and a macroscopic surface, regardless of the distance regime, as CP forces. Some authors refer to interactions at electrostatic distances as van der Waals forces.}. The distance to the medium, its electromagnetic response and its shape all affect Casimir-Polder forces, leading to a rich and varied effort to calculate them in diverse situations (see, e.g., \cite{mehlVanWaalsInteraction1980,marvinVanWaalsInteraction1982,wylieQuantumElectrodynamicsInterface1985,brevikCasimirPolderEffect1998,fordFocusingVacuumFluctuations2000,dalvitProbingQuantumVacuumGeometrical2008}).

It is important to note that dispersion forces -- including the CP force -- persist even when all sub-systems involved (electromagnetic field, atom/molecule) are in their ground states, making it difficult or impossible to `turn off' CP forces. The ever-progressing miniaturization of modern technology and the dramatic distance dependence of dispersion forces (anything from the first to \textit{eighth} inverse power of the distance, depending on the system at hand \cite{buhmannGroundstateVanWaals2005}), mean that they can no longer be ignored in technological settings.  They are especially important, for example, when an attractive force leads to stiction \cite{Stiction}, which is an stumbling block in the design of micro- and nano-electromechanical systems (MEMS/NEMS).  \cite{MEMSbook}.

\begin{figure}
 \begin{tikzpicture}

 \def\hWidth{1.5}
 \def\vHeight{2.3}
 \def\angle{0}
 \def\thickness{0.2}
 \def\dipoleSep{0.7}
 \def\dipoleLength{1}
 \def\dipolePosY{0.3}
 \def\reflectionOffset{0.2}
 \def\lineOffset{0.0}
 \def\atomScale{0.7}
  \def\arrowScale{0.2}
  \def\shiftH{2.5}
  \def\shiftV{3.5}
  \def\atomBodySep{1.2}
\node at (0.45,0.8*\vHeight)[anchor = north] {a) Plane};

\pic[scale = \atomScale] at (0.0,0) {atom}; 
\pic[scale = \arrowScale] at (0.8*\atomBodySep,0) {arrowRight}; 
\node[anchor = west ] at (-0.3,-0.5 ) {\tiny{Attraction}};

 
 \fill [Blue!20] (\atomBodySep,-0.5* \vHeight)rectangle (\atomBodySep+\thickness,0.5*\vHeight);
  \draw (\atomBodySep,-0.5*\vHeight) -- (\atomBodySep,0.5*\vHeight) ;

\tikzset{shift={(\shiftH,0)}}
\node at (0.85,0.8*\vHeight)[anchor = north] {b) Sphere/Cylinder};
 \filldraw [fill = Blue!20]  (\atomBodySep+0.7,0) circle (0.7);	
 \node[anchor = west ] at (-0.3,-0.5 ) {\tiny{Attraction}};
\pic[scale = \atomScale] at (0.0,0) {atom}; 
\pic[scale = \arrowScale] at (0.8*\atomBodySep,0) {arrowRight}; 
 
\tikzset{shift={(3.5,0)}}
\pic[scale = \atomScale] at (0.0,0) {atom}; 
\pic[scale = \arrowScale] at (0.8*\atomBodySep,0) {arrowRight}; 
\node[anchor = west ] at (-0.3,-0.5 ) {\tiny{Attraction}};
\node[text width = 3cm] at (1,0.8*\vHeight)[anchor = north] {c) General complex geometries};
 \pic at (\atomBodySep+1.5,0) {blob};

\tikzset{shift={(-2*\shiftH,0)}}

\tikzset{shift={(0,-\shiftV)}}
 \node[anchor = west ] at (-1.5,0.8*\vHeight) {d) Inverse design};
\pic[scale = \arrowScale] at (0.7-\arrowScale,0) {arrowLeft};
  \pic[scale = \atomScale] at (0.7,1) {atom}; 
    \pic[scale = 0.15] at (0.7,-1) {cylinder}; 
 \draw  (1.4,-1)rectangle node[text width = 3cm,align = center]{\footnotesize{Time-domain Casimir-Polder, \\ FDTD, \\ adjoint methods, \\ level-set advection.}} (4.2,1);
 \node[text width = 2.5cm] at (0,1 ) {\footnotesize{Atomic \\properties}};
\node[text width = 2.5cm] at (0,0 ) {\footnotesize{\underline{Goal} of\\ repulsion}};
\node[text width = 2.5cm] at (0,-1 ) {\footnotesize{Initial\\ (arbitrary)\\ geometry}};
\draw (1,-1.4) -- (1.2,-1.4) -- (1.2,1.4) -- (1,1.4);
\draw[->]  (1.2,0) -- (1.4,0);
\draw[->]  (4.2,0) -- (4.4,0);
\pic[scale = 0.5] at (6.2,0) {funnel};
\pic[scale = \atomScale] at (5.3,0) {atom}; 
\pic[scale = \arrowScale] at (4.7-\arrowScale,0) {arrowLeft};
 \node[anchor = west ] at (4.5,-0.5 ) {\tiny{Repulsion}};
 \draw (4.6,-1.4) -- ++(-0.2,0) -- ++ (0,2.8) -- ++(+0.2,0);
 \end{tikzpicture}
 \caption{Overview of the concept behind this work. All simple structures (and most complex ones) result in solely attractive forces on ground state atoms, as illustrated in the top row. Our approach to finding repulsive forces is based on specifying only the atomic properties, the goal of repulsion, and an arbitrary initial geometry. These are then fed into an algorithm relying on the ingredients listed in the figure, which will end up deforming the shape into one exhibiting Casimir-Polder repulsion.}\label{generalOverviewFig}
\end{figure}
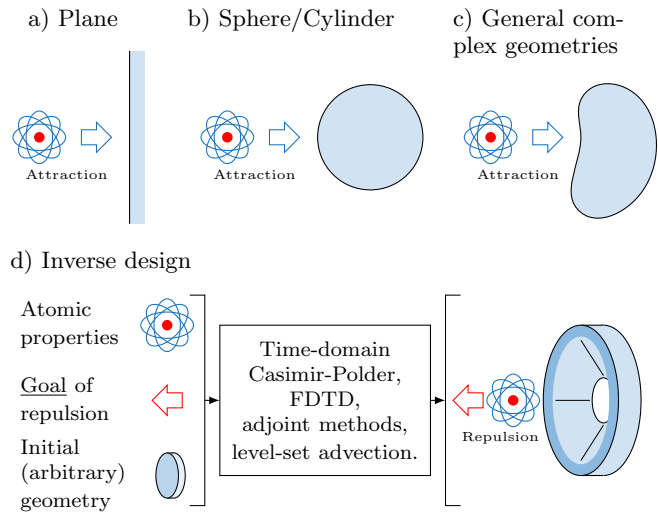

The ability to engineer repulsive dispersion forces is therefore of clear interest. This has proven somewhat difficult, however, with almost all physical setups resulting in attractive forces \cite{EquilCas}. It was pointed out as long ago as the 1970s that combinations of high permittivity and high permeability permit the occurrence of repulsive Casimir forces \cite{RepMag}, though this remains a theoretical exercise as any realistic material will have a permeability close to that of vacuum. More promisingly, dielectric slabs separated by an intervening, polarizable, dielectric fluid \cite{CPLiquid} (all of which can be non-magnetic) also show a repulsive force, but this concept is not transferrable to an atom in vacuum near a surface, which is the configuration of interest for CP forces. However, for polarizable atoms with sufficient anisotropy, it has been shown that repulsion in vacuum along a particular axis can occur near a wedge or plate-with-a-hole geometry \cite{miltonCasimirPolderRepulsionEdges2011,miltonCasimirPolderRepulsionPolarizable2012,miltonRepulsiveCasimirCasimir2012,Toroid,PWH}. The primary barrier to a more systematic approach to geometries supporting repulsion is that calculation of the Casimir-Polder force is computationally expensive for anything but the simplest geometries, largely because of the required integral over imaginary frequencies. This motivates us to look beyond combinations of elementary shapes and turn towards computer-aided design as schematically illustrated in Fig.~\ref{generalOverviewFig}.

In this Letter we present a method of producing a desired CP interaction via algorithmically created geometries. Our approach is general regarding the direction (sign) of the force that the structure exerts on a test atom, and also could be easily extended for excited atoms, but here we focus exclusively on the much more elusive case of repulsion of a ground-state atom. We want to clarify that we aim to position the structure ``in front of'' the atom, removing ambiguity in defining repulsive interaction, known from, e.g. ``zipper'' geometries of Ref. \cite{CasimirZip}. To help achieve this, we employ a powerful set of tools known as \textit{inverse design} that will allow us to algorithmically produce a structure with desired properties via a form of reverse-engineering. Initially developed in the context of mechanical \cite{ID_mech_eng} and aeronautical engineering \cite{IDAero1988}, inverse design has since expanded its range of applications and has established itself as a standard tool in nanophotonics \cite{IDReviewMolesky2018} and elsewhere (for recent advances see, e.g., 
\cite{wangMetasurfaceBasedSolidPoincare2023,senManyBodyQuantumInterference2024,cordaroSolvingIntegralEquations2023,bordigaAutomatedDiscoveryReprogrammable2024,goelInverseDesignHighdimensional2024}). The most relevant development in the context of CP forces is its application in light-matter interactions and macroscopic quantum electrodynamics (MQED) \cite{Rob_ID_MQED}, which has been applied to account for a range of fluctuation induced, atomic interactions with surfaces \cite{bennettInverseDesignEnvironmentinduced2021,matuszakShapeOptimizationsBodyassisted2022,capersDesigningCollectiveNonlocal2021,miguel-torcalInversedesignedDielectricCloaks2022a,ID_Rob_Claire}. The crucial ingredient in the method is that it takes advantage of the \textit{adjoint} method (for its application in electromagnetics problems, see Ref.~\cite{MillerPhD2012}), which allows for simultaneous optimization of an arbitrary number of features --- given a set of constraints --- using only two simulations per iteration. Its underlying mechanism exploits the source-observer symmetry of electromagnetic fields, which equivocates propagation of excitations from source to observer with their counterparts if the source and observer are swapped.

The CP potential $U_{\mathrm{CP}}$ of a ground-state atom or molecule at position $\mathbf{r}_{\mathrm{A}}$ is well-known (see, e.g., \cite{buhmann2012Book1}) to be expressible as $U_{\mathrm{CP}}(\mathbf{r}_{\mathrm{A}}) = -\frac{\hbar \mu_{0}}{2\pi}\mathrm{Im}\int_{0}^{\infty} \!\meas{\omega}\mathrm{Tr}\left[ \overline{\alpha}(\omega) \cdot \mathbb{G}^{(1)}(\mathbf{r}_{\mathrm{A}},\mathbf{r}_{\mathrm{A}},\omega)\right]$, where $\hbar$ is the reduced Planck's constant, $\mu_{0}$ is the vacuum permeability, $\overline{\alpha}(\omega)$ is the polarizability tensor. Finally, and most crucially,  $ \mathbb{G}^{(1)}(\mathbf{r},\mathbf{r}',\omega)$ is the (scattering part of the) dyadic Green's tensor that solves the Helmholtz equation for the geometry at hand, representing the likelihood of an excitation of frequency $\omega$ to travel from $\mathbf{r}'$ to $\mathbf{r}$, via the introduced structure. This Green's tensor holds all the information about the scattering properties of the system. Therefore, finding the $\mathbb{G}^{(1)}(\mathbf{r},\mathbf{r}',\omega)$ for a given $\omega$, and integrating over all frequencies weighted by the polarizability (generally a tensor) gives a result for the CP potential. The main difficulty in such calculations is that expressions for $\mathbb{G}^{(1)}$ are only known for the very simplest of geometries, such as infinite planes and perfect spheres. More complicated structures require either approximations (e.g., \cite{buhmannBornExpansionCasimirPolder2006}) or, more generally, a numerical approach. One obvious route towards this is to exploit the popularity and convenience offered by modern computational electromagnetic schemes such as finite-difference time-domain (FDTD), used extensively in photonics and industrial applications. In the form given above, $U_{\mathrm{CP}}$ is unsuitable for use with an FDTD solver, as each frequency $\omega$ would have to be evaluated separately. Hence, we turn to rewriting the CP interaction in the time domain, which emerged as a technique some time ago for Casimir-type forces \cite{Rodrig2007,Rodrig2009}. We will follow a recent version of this presented in Ref.~\cite{Francesco}, that extended time-domain approaches to include the treatment of the dispersive atomic polarizability via the discontinuous Galerkin time domain (DGTD) method. This approach efficiently computes the CP force, modeling complex geometries using an unstructured mesh. We will follow this recipe insofar as the time-domain formulation is concerned, but we will perform our field calculations on a standard grid system using the open-source FDTD solver MEEP \cite{Meep}.

Starting in the frequency picture, we consider a current source $\mathbf{J}(\mathbf{r},\omega) = \delta(\mathbf{r}-\mathbf{r}')J(\omega)\hat{\mathbf{e}}_{j}$, where $J(\omega)$ is the source frequency profile, and $\hat{\mathbf{e}}_{j}$ is a unit vector in the direction $j$. Since this is a point source, the $ij$-th component of the Green's tensor $\mathbb{G}(\mathbf{r},\mathbf{r}',\omega)$ and $E_{i}(\mathbf{r},\omega; \mathbf{r}',\hat{e}_{j})$ --- the $i$-th electric field component with its source pointing along $j$ --- are related through $\mathbb{G}_{ij}(\mathbf{r},\mathbf{r}',\omega) = -\mathrm{i} E_{i}(\mathbf{r},\omega; \mathbf{r}',\hat{e}_{j})/\omega \mu_{0}J(\omega)$. We assume the atom to be polarized in the $x$ direction, i.e. $\overline{\alpha}(\omega) = \alpha(\omega)\delta_{ix}$, simplifying the subsequent analysis. Correspondingly, we require only the diagonal Green's tensor components $\mathbb{G}_{ii}(\mathbf{r},\mathbf{r}',\omega)$, which, for our choice of polarizability, reduces to $\mathbb{G}_{xx}(\mathbf{r},\mathbf{r}',\omega)$, assembled via simulating electric fields from sources oscillating in the directions of the basis vectors. Next, following the approach adopted in \cite{Francesco}, we represent $U_{\mathrm{CP}}$ in the time-domain as a convolution
\begin{align}
\label{Utime}
U_{\mathrm{CP}}(\mathbf{r}_{\mathrm{A}})=-\hbar \int_{0}^{\infty} \meas{t}~ \mathrm{Im}[g_{x}(-t)]E^{(1)}_{x}(\mathbf{r}_{\mathrm{A}},t),
\end{align}
in which $E_{x}^{(1)}(\mathbf{r}_{\mathrm{A}},t)$ is the $x$ component of the scattered electric field at the coincidence limit ($\mathbf{r}' =\mathbf{r} =\mathbf{r}_{\mathrm{A}}$), and we define the function $g_{x}(t)$ as a Fourier transform of $g_{x}(\omega) = -\mathrm{i} \alpha(\omega)\frac{\omega}{J(\omega)}H(\omega)$, where $H(\omega)$ is the Heaviside step function enforcing causality. The source function is set to $J(t) = J_{0}\left[4(\gamma t)^{3}-(\gamma t)^{4} \right]e^{-\gamma t}H(t)$, in which $\gamma$ sets a cut-off time after which the current and the scattered fields will be damped substantially. This particular form is identical to that in \cite{Francesco}, which was selected as it satisfies the requirement that the fields should vanish at the start of the simulation, while having continuous first and second time derivatives at $t=0$. 

The tools we have discussed so far are enough to accurately calculate the CP potential landscape around an arbitrarily-shaped complex object, as discussed extensively in \cite{Francesco}. 
%
%
%
%
To find a shape capable of repulsion, our optimization goal boils down to maximizing the magnitude of the force  $\mathbf{F}_{\mathrm{CP}}(\mathbf{r}) = - \nabla U_{\mathrm{CP}}(\mathbf{r})$ in a specific direction ($x$, say), which we will choose as pointing away from a localized structure to be determined. We can define the time-dependent merit function (see the discussion in the appendix of \cite{MillerPhD2012}), responsible for the $x$- component of the force, as $ F_{x}[\mathbf{E}] = \int_{\mathcal{T}}\meas{t}~f_{x}[\mathbf{E}(\mathbf{r}_{\mathrm{A}},t)],$
where ${f_{x}[\mathbf{E}(\mathbf{r}_{\mathrm{A}},t)] \equiv \mathrm{Im}g(-t) \partial_{x}\sum_{i}E_{i}(\mathbf{r}_{\mathrm{A}},t).}$
The variation in the merit function is $\delta F_{x}=\int_{\mathcal{T}} \meas{t}~\delta f_{x} $ with
\begin{align}
\delta f_{x} & = -\hbar \int_{\mathcal{V}'}\!\measV{r}'\partial_{x''}\int_{\mathcal{T}'}\!\meas{t'}\notag \\
&\times\mathbf{P}(\mathbf{r}'',t')\cdot\mathbb{G}^{\mathrm{T}}(\mathbf{r}'',t',\mathbf{r}_{\mathrm{A}},t)\mathbf{j}(\mathbf{r}_{\mathrm{A}},t)\rvert_{\mathbf{r''}=\mathbf{r'}},
\end{align}
where $\mathcal{V'}$ is the transformed volume, $\mathrm{T}$ denotes the transpose, and $\mathbf{j}(\mathbf{r}_{\mathrm{A}},t)\equiv \mathrm{Im}g(-t)\hat{x}$ is the source's current vector. The ordering of the arguments $t$ and $t'$ of $\mathbb{G}^{\mathrm{T}}(\mathbf{r}'',t',\mathbf{r}_{\mathrm{A}},t)$ implies a curious property --- the measurement occurs \textit{before} the source gets excited, hinting at a reverse-time nature of this object, which is a feature that we will expand on shortly. Firstly, however, we focus on the defining property of the adjoint method; since the term associated with the Green's tensor in the integrand of $\delta f_{x}$ has the form of an electric field, we can treat it as such and thus define a fictitious, \textit{adjoint} field $\mathbf{E}^{\mathrm{A}}(\mathbf{r}',t') \equiv \int_{\mathcal{T}}\meas{t}~\mathbb{G}^{\mathrm{T}}(\mathbf{r}',t',\mathbf{r}_{\mathrm{A}},t)\mathbf{j}(\mathbf{r}_{\mathrm{A}}, t).$ We write the change in the merit function $F_{x}$ as 
\begin{equation}
\label{dF}
\delta F_{x} = -\int_{\mathcal{V'}}\measV{r}'\int_{\mathcal{T'}}\meas{t'}~ \partial_{x''}\mathbf{E}(\mathbf{r''},t')\cdot\mathbf{E}^{\mathrm{A}}(\mathbf{r''},t'),
\end{equation}
where the derivative is evaluated at $\mathbf{r''} = \mathbf{r'}$, and we made the assumption $\mathbf{P}(\mathbf{r},t)\propto \mathbf{E}(\mathbf{r},t)$ since we are only interested in the relative changes to the merit function, which also allowed us to drop the (positive) constants in front. We notice that the change in the merit function is dependent on only the overlap between the so-called \textit{forward} \footnote{The term forward is conventionally defined as opposite-to-adjoint, and \textit{not} as forward-time; however, this interpretation applies in this work.} field $\mathbf{E}(\mathbf{r},t)$, and the \textit{adjoint} field $\mathbf{E}^{\mathrm{A}}(\mathbf{r},t)$. Therefore, we need to compute only two simulations to determine the optimal change in our structure. The interchangeability of the source and observation points guaranteed in reciprocal media allows us to shift the spatial and temporal coordinates of the Green's tensor to the first two ``slots'', i.e. from $\mathbb{G}(\mathbf{r},t,\mathbf{r},t)$ to $\mathbb{G}^{\mathrm{T}}(\mathbf{r}',t',\mathbf{r},t)$. If this had not been the case (i.e., non-reciprocal structures), we would have had to have run a separate simulation for each position $\mathbf{r}$. This massive reduction in computational overhead is the advantage of the adjoint method, as discussed in detail in, for example, Ref.~\cite{IDReviewMolesky2018}. A major difference between this time-domain calculation and previous work is that ``swap'' affects spatial \emph{and} temporal coordinates, and the exchange of the latter necessitates computing $\mathbf{E}^{\mathrm{A}}(\mathbf{r},t)$ \emph{backwards} in time. 

\begin{figure}
    \centering
    \includegraphics[width=1\linewidth]{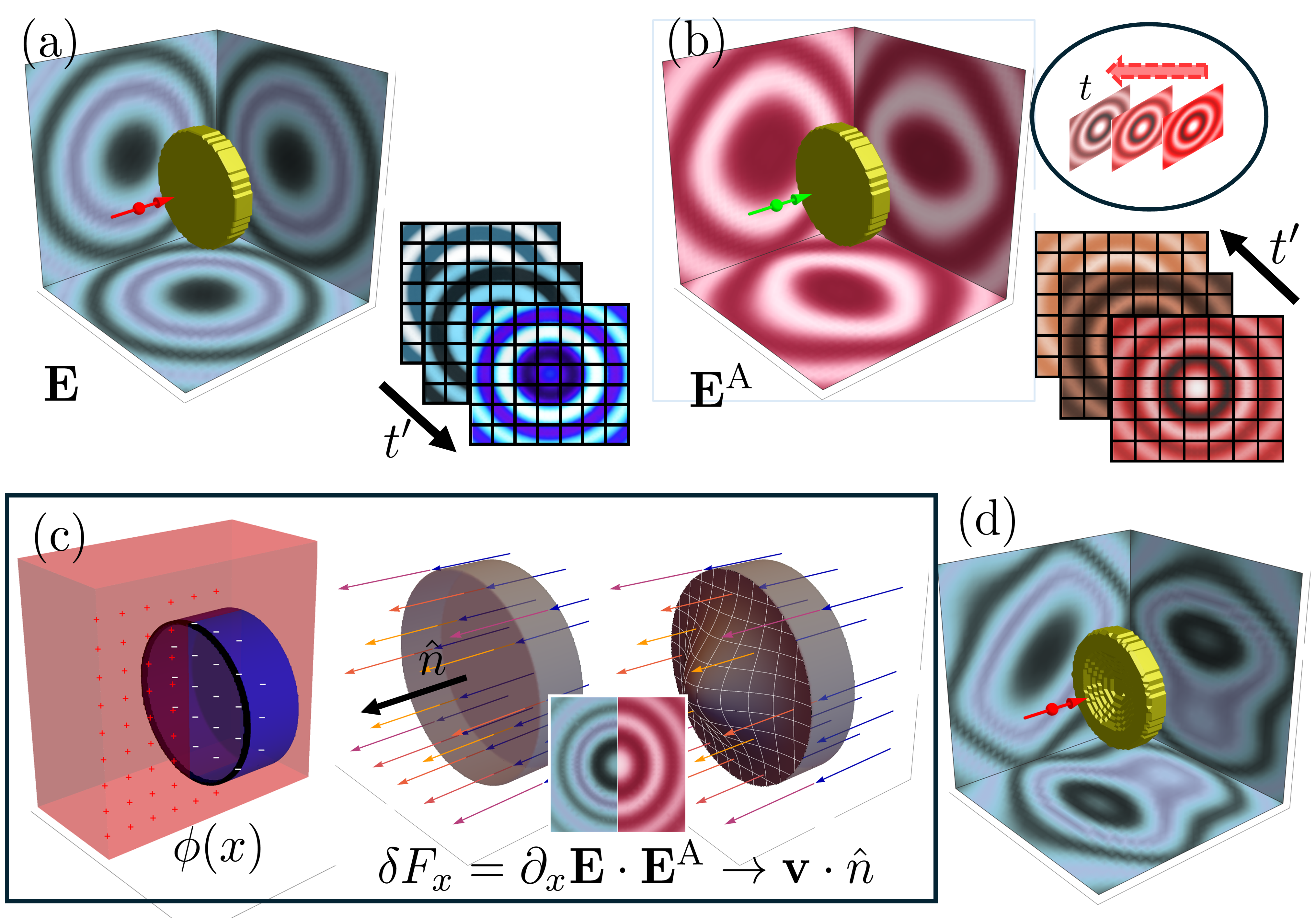}
    \caption{Overview of the algorithm. (a) An atomic dipole (in red) is placed in front of a gold cylinder, and $\mathbf{E}$ is recorded at a range of times $t'$ (shown as slices for clarity). (b) the adjoint dipole (in green) is excited at a range of times $t$, and the adjoint field $\mathbf{E}^{\mathrm{A}}$ is recorded along $t'$, and then the fields are integrated along $t$. Here, the time arrays are reversed, representing the adjoint field running backward in time. (c) We solve the advection equation with a `wind' velocity defined by the combination of forward and adjoint fields specified in the main text. This means the level-set boundaries ($\phi(x)=0$) incrementally migrate toward a new position that will be advantageous with respect to our merit function $F_{x}$, shown in (d). This in turn changes the forward and adjoint fields due to their having new boundary conditions, so the process repeats with the new geometry until the goal is reached.   }
    \label{fig:adjoint process}
\end{figure}

Equation \eqref{dF} tells us how the force changes after introducing a perturbation with volume $\mathcal{V}'$, but does not tell us how to introduce that perturbation in such a way that the force increases in the desired direction. We ensure this by taking advantage of the level-set formalism \cite{LevelSetOSHER}, operating on the principle of modifying and stretching an initial shape. The overall algorithm is illustrated in Fig.~\ref{fig:adjoint process}.
The shape is encoded through the position-dependent level set function $\Phi(\mathbf{r})$, with the zero contour $\Phi(\mathbf{r})= 0$ representing the shape's boundary, and consequently the regions $\Phi < 0$ and $\Phi >0$ constituting its interior and exterior, respectively. The level set function will be updated throughout the optimisation, to track this we define an artificial `time' variable $\tau$ (not to be confused with the physical times $t$ and $t'$), so that $\Phi(\mathbf{r})$ becomes $\Phi[\mathbf{r}(\tau)]$.
%
%
The propagation of fronts modeled by $\Phi[\mathbf{r}(\tau)]=0$ can be found by taking the total time derivative, resulting in the advection equation $\pder[\Phi]{\tau} + v_{n} |\nabla \Phi| = 0,$
where the $v_{n} = \mathbf{v}\cdot \hat{\mathbf{n}}$ is a (problem-specific) scalar field of the velocity along the direction normal to the boundary, denoted by $\hat{\mathbf{n}}$. We revisit Eq.~(\ref{dF}), modifying the integral measure, $\int_{\mathcal{V'}}\measV{r}'\rightarrow\int_{\partial \mathcal{V'}}\meas{A}v_{n}\delta\tau$, where we have incorporated an infinitesimal deformation in the perpendicular direction $\delta x'$, into the velocity term $v_{n} = \delta x'/\delta \tau$, and the integration is now over an area $\partial \mathcal{V'}$. Equation \eqref{dF} can now be rewritten as
\begin{align}
\delta F_{x} = -\int_{\partial\mathcal{V'}} \!\!\meas{A}v_{n}\delta\tau~ \partial_{x'}\int_{\mathcal{T'}}\!\meas{t'}~\mathbf{E}(\mathbf{r'},t')\cdot \mathbf{E}^{\mathrm{A}}(\mathbf{r'},t').
\end{align}
where, again, positive constants have been dropped. By setting $v_{n}= \partial_{x'}\int_{\mathcal{T'}}\meas{t'}~\mathbf{E}(\mathbf{r'},t')\cdot \mathbf{E}^{\mathrm{A}}(\mathbf{r'},t'),$ we achieve an expression for the variation in $F_{x}$ that does not change sign, i.e. $\delta F_{x} = -\int_{\partial \mathcal{V}}\meas{A}v_{n}^{2}\delta\tau$, making sure that $F_{x}$ evolves to be closer to the optimization goal with each iteration. By solving the advection equation with velocity $v_n$ (in this case using Python's PDE solver FIPY \cite{Fipy} and scikit-fmm \cite{scikit-fmm} to extend the velocity over the entire domain), in turn given by our simulated forward and adjoint fields, we drive the evolution of our shape's boundaries as prescribed by the particular velocity field.

We now have all the tools ready to proceed with the algorithmic discovery process, which we start by introducing a gold cylinder with its axis of symmetry pointing towards an non-isotropic, ground-state atom, placed in vacuum and polarised in the $x$-direction. As in Ref.~\cite{Francesco}, we choose the atom to be rubidium$-87$, having a dispersive polarizability of $\alpha(\omega) = \alpha_{0}\frac{\omega_{a}^{2}}{\omega_{a}^{2}-\omega^{2}-i \gamma_{a}\omega}$, in which $\omega_{a} = 1.6 \mathrm{eV}$ is the resonance frequency, $\gamma_{a} = 2.5\times 10 ^{-8} \mathrm{eV}$ is the linewidth of the dominating transition and $\alpha_{0}$ is the static polarizability. We choose an operational length scale $L_{0} = 100 \mathrm{nm}$, in terms of which we define the current cut-off point $\gamma = 2.5 c/L_{0}$.
We perform the CP energy calculations via simulated electric fields and run our adjoint optimization algorithm using MEEP. We use MEEP's material function based on a broadband response Drude model to represent the polarizability of gold. We define the simulation space as a cube of side length $8L_0$, surrounded on each side by a perfectly matched layer of depth $L_0$. The cylinder has radius $R=1.5L_0$ and height $h = 0.4L_0$, is coaxial with the $x$-axis, and positioned with its center at a distance $x_{0} = 1.1 L_0$ from the dipole.

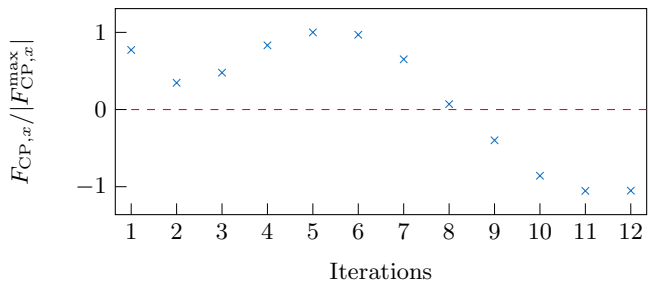
\begin{figure}
\begin{tikzpicture}
\begin{axis}[
width = \columnwidth,
height = 0.5*\columnwidth,
xlabel = {Iterations},
ylabel = {$F_{\mathrm{CP},x}/|{F}^{\mathrm{max}}_{\mathrm{CP},x}|$},                    
    minor tick num = 0,                
    tick style = {color=black},        
    enlargelimits = 0.15,               
    xtick pos=left, ytick pos=left,    
    xtick = {1,2,3,4,5,6,7,8,9,10,11,12},
    xmin = 2,
    xmax= 11,
    ]               
\addplot [only marks, Blue, point meta=y, mark=x] table [x=N, y=MF, col sep=comma] {data.csv};
\addplot [red, dashed] coordinates {(1, 0) (13, 0)};

\end{axis}
\end{tikzpicture}
\caption{The change in the normalized merit function ($F_{\mathrm{CP},x}/|{F}^{\mathrm{max}}_{\mathrm{CP},x}|$) over the course of the optimization.}
\label{fig:merit function}
\end{figure}

\begin{figure}
    \centering
    \includegraphics[width=0.8\linewidth]{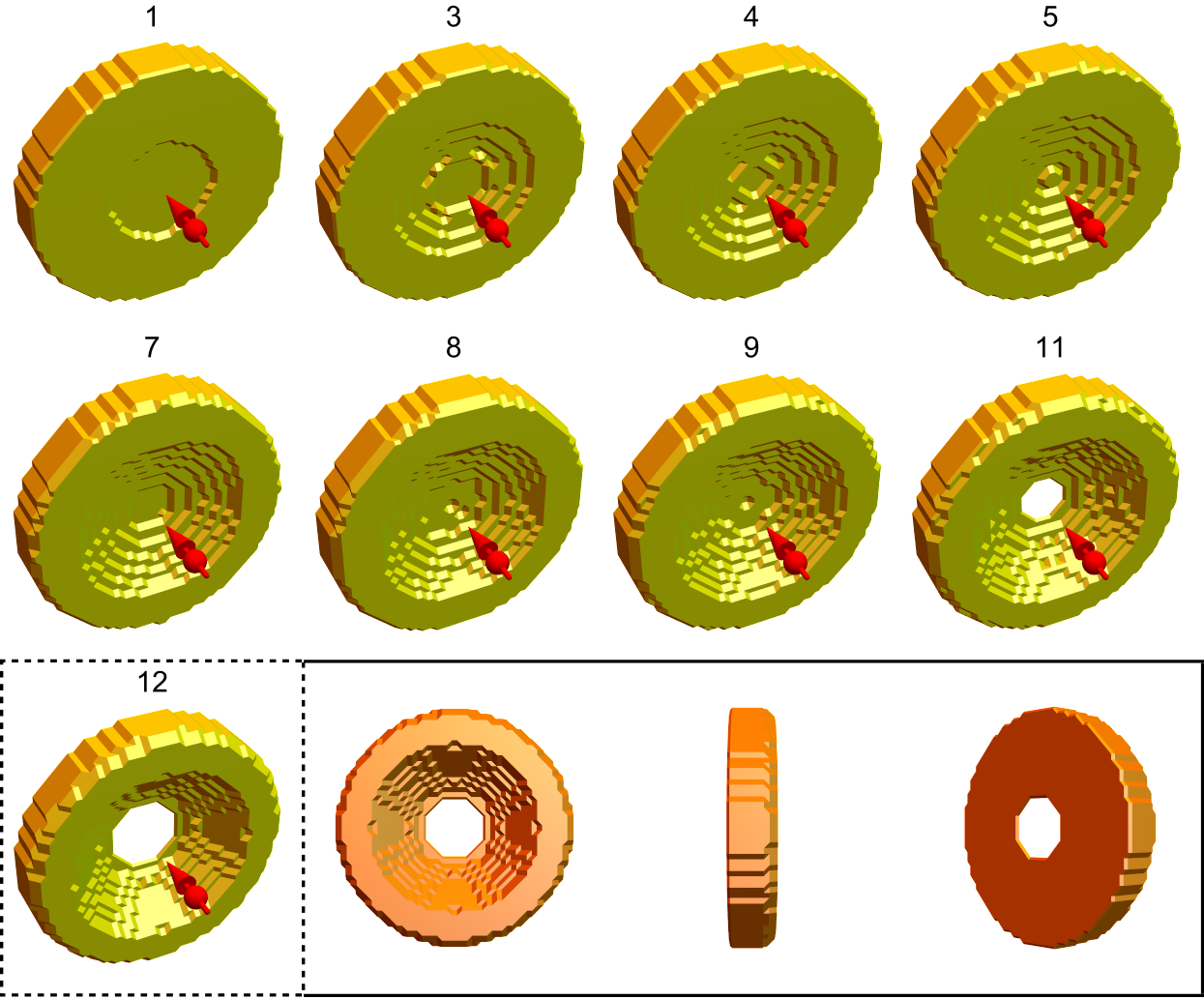}
    \caption{An $x$- polarised atom in front of an evolving structure (disk) which exerts progressively greater repulsive force. Here, the first shape corresponds to the state \textit{after} the first iteration. The dashed box in the bottom panel contains the structure after the final step, and the regular box in the last row shows the final structure (in a darker shade) from different viewpoints.}
    \label{fig:ls_evolution}
\end{figure}

After $12$ iterations, the method converges, and the value of the merit function settles to a negative (repulsive) value, as shown in Fig.~\ref{fig:merit function}.
The geometry to which the algorithm converges is a funnel-like shape, shown in the bottom left corner of Fig.~\ref{fig:ls_evolution}, which could be interpreted as its attempt at forging a ring.
Looking at the iterative changes occurring in the cylinder (disk), we can observe quite an intuitive sequence of events. Initially, the shape responds by forming a dent in the center, which deepens with every iteration --- up to the point where a hole is formed --- reminiscent of the electrostatic results of the plate with a hole or a torus \cite{miltonCasimirPolderRepulsionEdges2011,Toroid,PWH}. This confirms the intuitive understanding of electrostatic CP repulsion; the hole's presence forces a boundary condition at which the potential vanishes. This, coupled with the potential being zero at the other extreme, i.e., infinity, forces the gradient of the potential to change sign somewhere in the intermediate region. Our simulation results in a shape separated from the atom by the distance that falls in this intermediate region. The range of atom-surface separations at which a given initial shape will produce a repulsion is relatively short, i.e. $\lesssim \lambda/2$. However, a shape centered at some distance for which the merit function plateaus while the force was attractive will still produce a repulsion if it is moved to be within the gradient-changing region. It is essential to mention that if the initial cylinder --- or any solid, non-perforated structure --- is located too close to the atom (the exact distance depends on the overall dimensions, but $x \gtrsim \lambda/2$ is usually safe) the algorithm will not be able to overcome the attractive force to bring about repulsion. 


In this article we have presented a general algorithmic method of generating geometries that produce specific Casimir-Polder forces, demonstrating this for the compelling case of repulsive forces. The algorithm gives results reminiscent (but not identical) to their analytic counterparts, i.e., known shapes that exert a repulsive force on an anisotropic atom. Our method can serve as the basis not only for discovery of repulsive structures for more complicated geometries with real materials, but in any situation where control of the magnitude and direction of the Casimir-Polder force is important.    

\begin{acknowledgments}
\textit{Acknowledgments} -- It is a pleasure to acknowledge fruitful discussions with Francesco Intravaia and Bettina Beverungen. The authors also thank Salvatore Butera for use of computational resources. R.K. acknowledges financial support from the UK EPSRC Doctoral Training Programme grant EPSRC/DTP 2020/21/EP/T517896/1. R.B. and C.M.C. acknowledge financial support from UK EPSRC grant EP/W016486/1. 
\end{acknowledgments}

\bibliography{CP_ID,RB_bib}

\providecommand{\noopsort}[1]{}\providecommand{\singleletter}[1]{#1}%
\begin{thebibliography}{54}%
\makeatletter
\providecommand \@ifxundefined [1]{%
 \@ifx{#1\undefined}
}%
\providecommand \@ifnum [1]{%
 \ifnum #1\expandafter \@firstoftwo
 \else \expandafter \@secondoftwo
 \fi
}%
\providecommand \@ifx [1]{%
 \ifx #1\expandafter \@firstoftwo
 \else \expandafter \@secondoftwo
 \fi
}%
\providecommand \natexlab [1]{#1}%
\providecommand \enquote  [1]{``#1''}%
\providecommand \bibnamefont  [1]{#1}%
\providecommand \bibfnamefont [1]{#1}%
\providecommand \citenamefont [1]{#1}%
\providecommand \href@noop [0]{\@secondoftwo}%
\providecommand \href [0]{\begingroup \@sanitize@url \@href}%
\providecommand \@href[1]{\@@startlink{#1}\@@href}%
\providecommand \@@href[1]{\endgroup#1\@@endlink}%
\providecommand \@sanitize@url [0]{\catcode `\\12\catcode `\$12\catcode `\&12\catcode `\#12\catcode `\^12\catcode `\_12\catcode `\%12\relax}%
\providecommand \@@startlink[1]{}%
\providecommand \@@endlink[0]{}%
\providecommand \url  [0]{\begingroup\@sanitize@url \@url }%
\providecommand \@url [1]{\endgroup\@href {#1}{\urlprefix }}%
\providecommand \urlprefix  [0]{URL }%
\providecommand \Eprint [0]{\href }%
\providecommand \doibase [0]{https://doi.org/}%
\providecommand \selectlanguage [0]{\@gobble}%
\providecommand \bibinfo  [0]{\@secondoftwo}%
\providecommand \bibfield  [0]{\@secondoftwo}%
\providecommand \translation [1]{[#1]}%
\providecommand \BibitemOpen [0]{}%
\providecommand \bibitemStop [0]{}%
\providecommand \bibitemNoStop [0]{.\EOS\space}%
\providecommand \EOS [0]{\spacefactor3000\relax}%
\providecommand \BibitemShut  [1]{\csname bibitem#1\endcsname}%
\let\auto@bib@innerbib\@empty
\bibitem [{\citenamefont {Milonni}(2013)}]{milonni2013quantum}%
  \BibitemOpen
  \bibfield  {author} {\bibinfo {author} {\bibfnamefont {P.~W.}\ \bibnamefont {Milonni}},\ }\href@noop {} {\emph {\bibinfo {title} {The quantum vacuum: an introduction to quantum electrodynamics}}}\ (\bibinfo  {publisher} {Academic press},\ \bibinfo {year} {2013})\BibitemShut {NoStop}%
\bibitem [{\citenamefont {Casimir}(1948)}]{casimirAttractionTwoPerfectly1948}%
  \BibitemOpen
  \bibfield  {author} {\bibinfo {author} {\bibfnamefont {H.~B.~G.}\ \bibnamefont {Casimir}},\ }\href {https://doi.org/citeulike-article-id:8810715} {\bibfield  {journal} {\bibinfo  {journal} {Proc. K. Ned. Akad.}\ }\textbf {\bibinfo {volume} {360}},\ \bibinfo {pages} {793} (\bibinfo {year} {1948})}\BibitemShut {NoStop}%
\bibitem [{\citenamefont {Sparnaay}(1957)}]{sparnaayAttractiveForcesFlat1957}%
  \BibitemOpen
  \bibfield  {author} {\bibinfo {author} {\bibfnamefont {M.~J.}\ \bibnamefont {Sparnaay}},\ }\href {https://doi.org/10.1038/180334b0} {\bibfield  {journal} {\bibinfo  {journal} {Nature}\ }\textbf {\bibinfo {volume} {180}},\ \bibinfo {pages} {334} (\bibinfo {year} {1957})}\BibitemShut {NoStop}%
\bibitem [{\citenamefont {Lamoreaux}(1997)}]{lamoreauxDemonstrationCasimirForce1997}%
  \BibitemOpen
  \bibfield  {author} {\bibinfo {author} {\bibfnamefont {S.~K.}\ \bibnamefont {Lamoreaux}},\ }\href {https://doi.org/10.1103/PhysRevLett.78.5} {\bibfield  {journal} {\bibinfo  {journal} {Physical Review Letters}\ }\textbf {\bibinfo {volume} {78}},\ \bibinfo {pages} {5} (\bibinfo {year} {1997})}\BibitemShut {NoStop}%
\bibitem [{\citenamefont {Mohideen}\ and\ \citenamefont {Roy}(1998)}]{mohideenPrecisionMeasurementCasimir1998}%
  \BibitemOpen
  \bibfield  {author} {\bibinfo {author} {\bibfnamefont {U.}~\bibnamefont {Mohideen}}\ and\ \bibinfo {author} {\bibfnamefont {A.}~\bibnamefont {Roy}},\ }\href {https://doi.org/10.1103/PhysRevLett.81.4549} {\bibfield  {journal} {\bibinfo  {journal} {Physical Review Letters}\ }\textbf {\bibinfo {volume} {81}},\ \bibinfo {pages} {4549} (\bibinfo {year} {1998})}\BibitemShut {NoStop}%
\bibitem [{\citenamefont {Decca}\ \emph {et~al.}(2007)\citenamefont {Decca}, \citenamefont {L{\'o}pez}, \citenamefont {Fischbach}, \citenamefont {Klimchitskaya}, \citenamefont {Krause},\ and\ \citenamefont {Mostepanenko}}]{deccaTestsNewPhysics2007}%
  \BibitemOpen
  \bibfield  {author} {\bibinfo {author} {\bibfnamefont {R.~S.}\ \bibnamefont {Decca}}, \bibinfo {author} {\bibfnamefont {D.}~\bibnamefont {L{\'o}pez}}, \bibinfo {author} {\bibfnamefont {E.}~\bibnamefont {Fischbach}}, \bibinfo {author} {\bibfnamefont {G.~L.}\ \bibnamefont {Klimchitskaya}}, \bibinfo {author} {\bibfnamefont {D.~E.}\ \bibnamefont {Krause}},\ and\ \bibinfo {author} {\bibfnamefont {V.~M.}\ \bibnamefont {Mostepanenko}},\ }\href {https://doi.org/10.1103/PhysRevD.75.077101} {\bibfield  {journal} {\bibinfo  {journal} {Physical Review D}\ }\textbf {\bibinfo {volume} {75}},\ \bibinfo {pages} {077101} (\bibinfo {year} {2007})}\BibitemShut {NoStop}%
\bibitem [{\citenamefont {Munday}\ \emph {et~al.}(2009{\natexlab{a}})\citenamefont {Munday}, \citenamefont {Capasso},\ and\ \citenamefont {Parsegian}}]{mundayMeasuredLongrangeRepulsive2009}%
  \BibitemOpen
  \bibfield  {author} {\bibinfo {author} {\bibfnamefont {J.~N.}\ \bibnamefont {Munday}}, \bibinfo {author} {\bibfnamefont {F.}~\bibnamefont {Capasso}},\ and\ \bibinfo {author} {\bibfnamefont {V.~A.}\ \bibnamefont {Parsegian}},\ }\href {https://doi.org/10.1038/nature07610} {\bibfield  {journal} {\bibinfo  {journal} {Nature}\ }\textbf {\bibinfo {volume} {457}},\ \bibinfo {pages} {170} (\bibinfo {year} {2009}{\natexlab{a}})}\BibitemShut {NoStop}%
\bibitem [{\citenamefont {Garrett}\ \emph {et~al.}(2018)\citenamefont {Garrett}, \citenamefont {Somers},\ and\ \citenamefont {Munday}}]{garrettMeasurementCasimirForce2018}%
  \BibitemOpen
  \bibfield  {author} {\bibinfo {author} {\bibfnamefont {J.~L.}\ \bibnamefont {Garrett}}, \bibinfo {author} {\bibfnamefont {D.~A.}\ \bibnamefont {Somers}},\ and\ \bibinfo {author} {\bibfnamefont {J.~N.}\ \bibnamefont {Munday}},\ }\href {https://doi.org/10.1103/PhysRevLett.120.040401} {\bibfield  {journal} {\bibinfo  {journal} {Physical Review Letters}\ }\textbf {\bibinfo {volume} {120}},\ \bibinfo {pages} {040401} (\bibinfo {year} {2018})}\BibitemShut {NoStop}%
\bibitem [{\citenamefont {Lifshitz}(1956)}]{Lifshitz1956}%
  \BibitemOpen
  \bibfield  {author} {\bibinfo {author} {\bibfnamefont {E.~M.}\ \bibnamefont {Lifshitz}},\ }\href@noop {} {\bibfield  {journal} {\bibinfo  {journal} {Zh. Eksp. Teor. Fiz.}\ }\textbf {\bibinfo {volume} {29}},\ \bibinfo {pages} {94} (\bibinfo {year} {1956})},\ \bibinfo {note} {[Sov. Phys. JETP. 2, 73 (1956)]}\BibitemShut {NoStop}%
\bibitem [{\citenamefont {I.E.~Dzyaloshinskii}\ and\ \citenamefont {Pitaevskii}(1961)}]{Lifshitz}%
  \BibitemOpen
  \bibfield  {author} {\bibinfo {author} {\bibfnamefont {E.~L.}\ \bibnamefont {I.E.~Dzyaloshinskii}}\ and\ \bibinfo {author} {\bibfnamefont {L.}~\bibnamefont {Pitaevskii}},\ }\href {https://doi.org/10.1080/00018736100101281} {\bibfield  {journal} {\bibinfo  {journal} {Advances in Physics}\ }\textbf {\bibinfo {volume} {10}},\ \bibinfo {pages} {165} (\bibinfo {year} {1961})}\BibitemShut {NoStop}%
\bibitem [{\citenamefont {Buhmann}(2012)}]{buhmann2012Book1}%
  \BibitemOpen
  \bibfield  {author} {\bibinfo {author} {\bibfnamefont {S.~Y.}\ \bibnamefont {Buhmann}},\ }\href {https://doi.org/10.1007/978-3-642-32484-0} {\emph {\bibinfo {title} {Dispersion {{Forces I}} - {{Macroscopic Quantum Electrodynamics}} and Ground-State {{Casimir}}, {{Casimir-Polder}} and van Der {{Waals Forces}}}}},\ Vol.\ \bibinfo {volume} {247}\ (\bibinfo  {publisher} {Springer},\ \bibinfo {address} {Berlin},\ \bibinfo {year} {2012})\BibitemShut {NoStop}%
\bibitem [{Note1()}]{Note1}%
  \BibitemOpen
  \bibinfo {note} {It is crucial for us to point out that we refer to any interaction between a polarizable atom and a macroscopic surface, regardless of the distance regime, as CP forces. Some authors refer to interactions at electrostatic distances as van der Waals forces.}\BibitemShut {Stop}%
\bibitem [{\citenamefont {Mehl}\ and\ \citenamefont {Schaich}(1980)}]{mehlVanWaalsInteraction1980}%
  \BibitemOpen
  \bibfield  {author} {\bibinfo {author} {\bibfnamefont {M.~J.}\ \bibnamefont {Mehl}}\ and\ \bibinfo {author} {\bibfnamefont {W.~L.}\ \bibnamefont {Schaich}},\ }\href {https://doi.org/10.1016/0039-6028(80)90553-1} {\bibfield  {journal} {\bibinfo  {journal} {Surface Science}\ }\textbf {\bibinfo {volume} {99}},\ \bibinfo {pages} {553} (\bibinfo {year} {1980})}\BibitemShut {NoStop}%
\bibitem [{\citenamefont {Marvin}\ and\ \citenamefont {Toigo}(1982)}]{marvinVanWaalsInteraction1982}%
  \BibitemOpen
  \bibfield  {author} {\bibinfo {author} {\bibfnamefont {A.~M.}\ \bibnamefont {Marvin}}\ and\ \bibinfo {author} {\bibfnamefont {F.}~\bibnamefont {Toigo}},\ }\href {https://doi.org/10.1103/PhysRevA.25.782} {\bibfield  {journal} {\bibinfo  {journal} {Physical Review A}\ }\textbf {\bibinfo {volume} {25}},\ \bibinfo {pages} {782} (\bibinfo {year} {1982})}\BibitemShut {NoStop}%
\bibitem [{\citenamefont {Wylie}\ and\ \citenamefont {Sipe}(1985)}]{wylieQuantumElectrodynamicsInterface1985}%
  \BibitemOpen
  \bibfield  {author} {\bibinfo {author} {\bibfnamefont {J.~M.}\ \bibnamefont {Wylie}}\ and\ \bibinfo {author} {\bibfnamefont {J.~E.}\ \bibnamefont {Sipe}},\ }\href {https://doi.org/10.1103/PhysRevA.32.2030} {\bibfield  {journal} {\bibinfo  {journal} {Physical Review A}\ }\textbf {\bibinfo {volume} {32}},\ \bibinfo {pages} {2030} (\bibinfo {year} {1985})}\BibitemShut {NoStop}%
\bibitem [{\citenamefont {Brevik}\ \emph {et~al.}(1998)\citenamefont {Brevik}, \citenamefont {Lygren},\ and\ \citenamefont {Marachevsky}}]{brevikCasimirPolderEffect1998}%
  \BibitemOpen
  \bibfield  {author} {\bibinfo {author} {\bibfnamefont {I.}~\bibnamefont {Brevik}}, \bibinfo {author} {\bibfnamefont {M.}~\bibnamefont {Lygren}},\ and\ \bibinfo {author} {\bibfnamefont {V.~N.}\ \bibnamefont {Marachevsky}},\ }\href {https://doi.org/10.1006/aphy.1998.5814} {\bibfield  {journal} {\bibinfo  {journal} {Annals of Physics}\ }\textbf {\bibinfo {volume} {267}},\ \bibinfo {pages} {134} (\bibinfo {year} {1998})}\BibitemShut {NoStop}%
\bibitem [{\citenamefont {Ford}\ and\ \citenamefont {Svaiter}(2000)}]{fordFocusingVacuumFluctuations2000}%
  \BibitemOpen
  \bibfield  {author} {\bibinfo {author} {\bibfnamefont {L.~H.}\ \bibnamefont {Ford}}\ and\ \bibinfo {author} {\bibfnamefont {N.~F.}\ \bibnamefont {Svaiter}},\ }\href {https://doi.org/10.1103/PhysRevA.62.062105} {\bibfield  {journal} {\bibinfo  {journal} {Physical Review A}\ }\textbf {\bibinfo {volume} {62}},\ \bibinfo {pages} {062105} (\bibinfo {year} {2000})}\BibitemShut {NoStop}%
\bibitem [{\citenamefont {Dalvit}\ \emph {et~al.}(2008)\citenamefont {Dalvit}, \citenamefont {Maia~Neto}, \citenamefont {Lambrecht},\ and\ \citenamefont {Reynaud}}]{dalvitProbingQuantumVacuumGeometrical2008}%
  \BibitemOpen
  \bibfield  {author} {\bibinfo {author} {\bibfnamefont {D.~A.~R.}\ \bibnamefont {Dalvit}}, \bibinfo {author} {\bibfnamefont {P.~A.}\ \bibnamefont {Maia~Neto}}, \bibinfo {author} {\bibfnamefont {A.}~\bibnamefont {Lambrecht}},\ and\ \bibinfo {author} {\bibfnamefont {S.}~\bibnamefont {Reynaud}},\ }\href {https://doi.org/10.1103/PhysRevLett.100.040405} {\bibfield  {journal} {\bibinfo  {journal} {Physical Review Letters}\ }\textbf {\bibinfo {volume} {100}},\ \bibinfo {pages} {040405} (\bibinfo {year} {2008})}\BibitemShut {NoStop}%
\bibitem [{\citenamefont {Buhmann}\ \emph {et~al.}(2005)\citenamefont {Buhmann}, \citenamefont {Welsch},\ and\ \citenamefont {Kampf}}]{buhmannGroundstateVanWaals2005}%
  \BibitemOpen
  \bibfield  {author} {\bibinfo {author} {\bibfnamefont {S.~Y.}\ \bibnamefont {Buhmann}}, \bibinfo {author} {\bibfnamefont {D.-G.}\ \bibnamefont {Welsch}},\ and\ \bibinfo {author} {\bibfnamefont {T.}~\bibnamefont {Kampf}},\ }\href {https://doi.org/10.1103/PhysRevA.72.032112} {\bibfield  {journal} {\bibinfo  {journal} {Physical Review A}\ }\textbf {\bibinfo {volume} {72}},\ \bibinfo {pages} {032112} (\bibinfo {year} {2005})}\BibitemShut {NoStop}%
\bibitem [{\citenamefont {Buks}\ and\ \citenamefont {Roukes}(2001)}]{Stiction}%
  \BibitemOpen
  \bibfield  {author} {\bibinfo {author} {\bibfnamefont {E.}~\bibnamefont {Buks}}\ and\ \bibinfo {author} {\bibfnamefont {M.~L.}\ \bibnamefont {Roukes}},\ }\href {https://doi.org/10.1103/PhysRevB.63.033402} {\bibfield  {journal} {\bibinfo  {journal} {Phys. Rev. B}\ }\textbf {\bibinfo {volume} {63}},\ \bibinfo {pages} {033402} (\bibinfo {year} {2001})}\BibitemShut {NoStop}%
\bibitem [{\citenamefont {Maluf}\ and\ \citenamefont {Williams}(2004)}]{MEMSbook}%
  \BibitemOpen
  \bibfield  {author} {\bibinfo {author} {\bibfnamefont {N.}~\bibnamefont {Maluf}}\ and\ \bibinfo {author} {\bibfnamefont {K.}~\bibnamefont {Williams}},\ }\href {https://books.google.co.uk/books?id=20j7IaDKlOUC} {\emph {\bibinfo {title} {An Introduction to Microelectromechanical Systems Engineering}}},\ Artech House Microelectromechanical Systems\ (\bibinfo  {publisher} {Artech House},\ \bibinfo {year} {2004})\BibitemShut {NoStop}%
\bibitem [{\citenamefont {Rahi}\ \emph {et~al.}(2010)\citenamefont {Rahi}, \citenamefont {Kardar},\ and\ \citenamefont {Emig}}]{EquilCas}%
  \BibitemOpen
  \bibfield  {author} {\bibinfo {author} {\bibfnamefont {S.~J.}\ \bibnamefont {Rahi}}, \bibinfo {author} {\bibfnamefont {M.}~\bibnamefont {Kardar}},\ and\ \bibinfo {author} {\bibfnamefont {T.}~\bibnamefont {Emig}},\ }\href {https://doi.org/10.1103/PhysRevLett.105.070404} {\bibfield  {journal} {\bibinfo  {journal} {Phys. Rev. Lett.}\ }\textbf {\bibinfo {volume} {105}},\ \bibinfo {pages} {070404} (\bibinfo {year} {2010})}\BibitemShut {NoStop}%
\bibitem [{\citenamefont {Boyer}(1974)}]{RepMag}%
  \BibitemOpen
  \bibfield  {author} {\bibinfo {author} {\bibfnamefont {T.~H.}\ \bibnamefont {Boyer}},\ }\href {https://doi.org/10.1103/PhysRevA.9.2078} {\bibfield  {journal} {\bibinfo  {journal} {Phys. Rev. A}\ }\textbf {\bibinfo {volume} {9}},\ \bibinfo {pages} {2078} (\bibinfo {year} {1974})}\BibitemShut {NoStop}%
\bibitem [{\citenamefont {Munday}\ \emph {et~al.}(2009{\natexlab{b}})\citenamefont {Munday}, \citenamefont {Capasso},\ and\ \citenamefont {Parsegian}}]{CPLiquid}%
  \BibitemOpen
  \bibfield  {author} {\bibinfo {author} {\bibfnamefont {J.~N.}\ \bibnamefont {Munday}}, \bibinfo {author} {\bibfnamefont {F.}~\bibnamefont {Capasso}},\ and\ \bibinfo {author} {\bibfnamefont {V.~A.}\ \bibnamefont {Parsegian}},\ }\href {https://doi.org/10.1038/nature07610} {\bibfield  {journal} {\bibinfo  {journal} {Nature}\ }\textbf {\bibinfo {volume} {457}},\ \bibinfo {pages} {170} (\bibinfo {year} {2009}{\natexlab{b}})}\BibitemShut {NoStop}%
\bibitem [{\citenamefont {Milton}\ \emph {et~al.}(2011)\citenamefont {Milton}, \citenamefont {Abalo}, \citenamefont {Parashar}, \citenamefont {Pourtolami}, \citenamefont {Brevik},\ and\ \citenamefont {Ellingsen}}]{miltonCasimirPolderRepulsionEdges2011}%
  \BibitemOpen
  \bibfield  {author} {\bibinfo {author} {\bibfnamefont {K.~A.}\ \bibnamefont {Milton}}, \bibinfo {author} {\bibfnamefont {E.~K.}\ \bibnamefont {Abalo}}, \bibinfo {author} {\bibfnamefont {P.}~\bibnamefont {Parashar}}, \bibinfo {author} {\bibfnamefont {N.}~\bibnamefont {Pourtolami}}, \bibinfo {author} {\bibfnamefont {I.}~\bibnamefont {Brevik}},\ and\ \bibinfo {author} {\bibfnamefont {S.~{\AA}.}\ \bibnamefont {Ellingsen}},\ }\href {https://doi.org/10.1103/PhysRevA.83.062507} {\bibfield  {journal} {\bibinfo  {journal} {Physical Review A}\ }\textbf {\bibinfo {volume} {83}},\ \bibinfo {pages} {062507} (\bibinfo {year} {2011})}\BibitemShut {NoStop}%
\bibitem [{\citenamefont {Milton}\ \emph {et~al.}(2012{\natexlab{a}})\citenamefont {Milton}, \citenamefont {Parashar}, \citenamefont {Pourtolami},\ and\ \citenamefont {Brevik}}]{miltonCasimirPolderRepulsionPolarizable2012}%
  \BibitemOpen
  \bibfield  {author} {\bibinfo {author} {\bibfnamefont {K.~A.}\ \bibnamefont {Milton}}, \bibinfo {author} {\bibfnamefont {P.}~\bibnamefont {Parashar}}, \bibinfo {author} {\bibfnamefont {N.}~\bibnamefont {Pourtolami}},\ and\ \bibinfo {author} {\bibfnamefont {I.}~\bibnamefont {Brevik}},\ }\href {https://doi.org/10.1103/PhysRevD.85.025008} {\bibfield  {journal} {\bibinfo  {journal} {Physical Review D}\ }\textbf {\bibinfo {volume} {85}},\ \bibinfo {pages} {025008} (\bibinfo {year} {2012}{\natexlab{a}})}\BibitemShut {NoStop}%
\bibitem [{\citenamefont {Milton}\ \emph {et~al.}(2012{\natexlab{b}})\citenamefont {Milton}, \citenamefont {Abalo}, \citenamefont {Parashar}, \citenamefont {Pourtolami}, \citenamefont {Brevik},\ and\ \citenamefont {Ellingsen}}]{miltonRepulsiveCasimirCasimir2012}%
  \BibitemOpen
  \bibfield  {author} {\bibinfo {author} {\bibfnamefont {K.~A.}\ \bibnamefont {Milton}}, \bibinfo {author} {\bibfnamefont {E.~K.}\ \bibnamefont {Abalo}}, \bibinfo {author} {\bibfnamefont {P.}~\bibnamefont {Parashar}}, \bibinfo {author} {\bibfnamefont {N.}~\bibnamefont {Pourtolami}}, \bibinfo {author} {\bibfnamefont {I.}~\bibnamefont {Brevik}},\ and\ \bibinfo {author} {\bibfnamefont {S.~{\AA}.}\ \bibnamefont {Ellingsen}},\ }\href {https://doi.org/10.1088/1751-8113/45/37/374006} {\bibfield  {journal} {\bibinfo  {journal} {Journal of Physics A: Mathematical and Theoretical}\ }\textbf {\bibinfo {volume} {45}},\ \bibinfo {pages} {374006} (\bibinfo {year} {2012}{\natexlab{b}})}\BibitemShut {NoStop}%
\bibitem [{\citenamefont {Abrantes}\ \emph {et~al.}(2018)\citenamefont {Abrantes}, \citenamefont {Fran\ifmmode~\mbox{\c{c}}\else \c{c}\fi{}a}, \citenamefont {da~Rosa}, \citenamefont {Farina},\ and\ \citenamefont {de~Melo~e Souza}}]{Toroid}%
  \BibitemOpen
  \bibfield  {author} {\bibinfo {author} {\bibfnamefont {P.~P.}\ \bibnamefont {Abrantes}}, \bibinfo {author} {\bibfnamefont {Y.}~\bibnamefont {Fran\ifmmode~\mbox{\c{c}}\else \c{c}\fi{}a}}, \bibinfo {author} {\bibfnamefont {F.~S.~S.}\ \bibnamefont {da~Rosa}}, \bibinfo {author} {\bibfnamefont {C.}~\bibnamefont {Farina}},\ and\ \bibinfo {author} {\bibfnamefont {R.}~\bibnamefont {de~Melo~e Souza}},\ }\href {https://doi.org/10.1103/PhysRevA.98.012511} {\bibfield  {journal} {\bibinfo  {journal} {Phys. Rev. A}\ }\textbf {\bibinfo {volume} {98}},\ \bibinfo {pages} {012511} (\bibinfo {year} {2018})}\BibitemShut {NoStop}%
\bibitem [{\citenamefont {Eberlein}\ and\ \citenamefont {Zietal}(2011)}]{PWH}%
  \BibitemOpen
  \bibfield  {author} {\bibinfo {author} {\bibfnamefont {C.}~\bibnamefont {Eberlein}}\ and\ \bibinfo {author} {\bibfnamefont {R.}~\bibnamefont {Zietal}},\ }\href {https://doi.org/10.1103/PhysRevA.83.052514} {\bibfield  {journal} {\bibinfo  {journal} {Phys. Rev. A}\ }\textbf {\bibinfo {volume} {83}},\ \bibinfo {pages} {052514} (\bibinfo {year} {2011})}\BibitemShut {NoStop}%
\bibitem [{\citenamefont {Rodriguez}\ \emph {et~al.}(2008)\citenamefont {Rodriguez}, \citenamefont {Joannopoulos},\ and\ \citenamefont {Johnson}}]{CasimirZip}%
  \BibitemOpen
  \bibfield  {author} {\bibinfo {author} {\bibfnamefont {A.~W.}\ \bibnamefont {Rodriguez}}, \bibinfo {author} {\bibfnamefont {J.~D.}\ \bibnamefont {Joannopoulos}},\ and\ \bibinfo {author} {\bibfnamefont {S.~G.}\ \bibnamefont {Johnson}},\ }\href {https://doi.org/10.1103/PhysRevA.77.062107} {\bibfield  {journal} {\bibinfo  {journal} {Phys. Rev. A}\ }\textbf {\bibinfo {volume} {77}},\ \bibinfo {pages} {062107} (\bibinfo {year} {2008})}\BibitemShut {NoStop}%
\bibitem [{\citenamefont {Bendsøe}\ and\ \citenamefont {Kikuchi}(1988)}]{ID_mech_eng}%
  \BibitemOpen
  \bibfield  {author} {\bibinfo {author} {\bibfnamefont {M.~P.}\ \bibnamefont {Bendsøe}}\ and\ \bibinfo {author} {\bibfnamefont {N.}~\bibnamefont {Kikuchi}},\ }\href {https://doi.org/https://doi.org/10.1016/0045-7825(88)90086-2} {\bibfield  {journal} {\bibinfo  {journal} {Computer Methods in Applied Mechanics and Engineering}\ }\textbf {\bibinfo {volume} {71}},\ \bibinfo {pages} {197} (\bibinfo {year} {1988})}\BibitemShut {NoStop}%
\bibitem [{\citenamefont {Jameson}(1988)}]{IDAero1988}%
  \BibitemOpen
  \bibfield  {author} {\bibinfo {author} {\bibfnamefont {A.}~\bibnamefont {Jameson}},\ }\href {https://doi.org/10.1007/BF01061285} {\bibfield  {journal} {\bibinfo  {journal} {Journal of Scientific Computing}\ }\textbf {\bibinfo {volume} {3}},\ \bibinfo {pages} {233} (\bibinfo {year} {1988})}\BibitemShut {NoStop}%
\bibitem [{\citenamefont {Molesky}\ \emph {et~al.}(2018)\citenamefont {Molesky}, \citenamefont {Lin}, \citenamefont {Piggott}, \citenamefont {Jin}, \citenamefont {Vucković},\ and\ \citenamefont {Rodriguez}}]{IDReviewMolesky2018}%
  \BibitemOpen
  \bibfield  {author} {\bibinfo {author} {\bibfnamefont {S.}~\bibnamefont {Molesky}}, \bibinfo {author} {\bibfnamefont {Z.}~\bibnamefont {Lin}}, \bibinfo {author} {\bibfnamefont {A.~Y.}\ \bibnamefont {Piggott}}, \bibinfo {author} {\bibfnamefont {W.}~\bibnamefont {Jin}}, \bibinfo {author} {\bibfnamefont {J.}~\bibnamefont {Vucković}},\ and\ \bibinfo {author} {\bibfnamefont {A.~W.}\ \bibnamefont {Rodriguez}},\ }\href {https://doi.org/10.1038/s41566-018-0246-9} {\bibfield  {journal} {\bibinfo  {journal} {Nature Photonics}\ }\textbf {\bibinfo {volume} {12}},\ \bibinfo {pages} {659} (\bibinfo {year} {2018})}\BibitemShut {NoStop}%
\bibitem [{\citenamefont {Wang}\ \emph {et~al.}(2023)\citenamefont {Wang}, \citenamefont {Wen}, \citenamefont {Deng}, \citenamefont {Li},\ and\ \citenamefont {Yang}}]{wangMetasurfaceBasedSolidPoincare2023}%
  \BibitemOpen
  \bibfield  {author} {\bibinfo {author} {\bibfnamefont {S.}~\bibnamefont {Wang}}, \bibinfo {author} {\bibfnamefont {S.}~\bibnamefont {Wen}}, \bibinfo {author} {\bibfnamefont {Z.-L.}\ \bibnamefont {Deng}}, \bibinfo {author} {\bibfnamefont {X.}~\bibnamefont {Li}},\ and\ \bibinfo {author} {\bibfnamefont {Y.}~\bibnamefont {Yang}},\ }\href {https://doi.org/10.1103/PhysRevLett.130.123801} {\bibfield  {journal} {\bibinfo  {journal} {Physical Review Letters}\ }\textbf {\bibinfo {volume} {130}},\ \bibinfo {pages} {123801} (\bibinfo {year} {2023})}\BibitemShut {NoStop}%
\bibitem [{\citenamefont {Sen}\ and\ \citenamefont {Mitchell}(2024)}]{senManyBodyQuantumInterference2024}%
  \BibitemOpen
  \bibfield  {author} {\bibinfo {author} {\bibfnamefont {S.}~\bibnamefont {Sen}}\ and\ \bibinfo {author} {\bibfnamefont {A.~K.}\ \bibnamefont {Mitchell}},\ }\href {https://doi.org/10.1103/PhysRevLett.133.076501} {\bibfield  {journal} {\bibinfo  {journal} {Physical Review Letters}\ }\textbf {\bibinfo {volume} {133}},\ \bibinfo {pages} {076501} (\bibinfo {year} {2024})}\BibitemShut {NoStop}%
\bibitem [{\citenamefont {Cordaro}\ \emph {et~al.}(2023)\citenamefont {Cordaro}, \citenamefont {Edwards}, \citenamefont {Nikkhah}, \citenamefont {Al{\`u}}, \citenamefont {Engheta},\ and\ \citenamefont {Polman}}]{cordaroSolvingIntegralEquations2023}%
  \BibitemOpen
  \bibfield  {author} {\bibinfo {author} {\bibfnamefont {A.}~\bibnamefont {Cordaro}}, \bibinfo {author} {\bibfnamefont {B.}~\bibnamefont {Edwards}}, \bibinfo {author} {\bibfnamefont {V.}~\bibnamefont {Nikkhah}}, \bibinfo {author} {\bibfnamefont {A.}~\bibnamefont {Al{\`u}}}, \bibinfo {author} {\bibfnamefont {N.}~\bibnamefont {Engheta}},\ and\ \bibinfo {author} {\bibfnamefont {A.}~\bibnamefont {Polman}},\ }\href {https://doi.org/10.1038/s41565-022-01297-9} {\bibfield  {journal} {\bibinfo  {journal} {Nature Nanotechnology}\ }\textbf {\bibinfo {volume} {18}},\ \bibinfo {pages} {365} (\bibinfo {year} {2023})}\BibitemShut {NoStop}%
\bibitem [{\citenamefont {Bordiga}\ \emph {et~al.}(2024)\citenamefont {Bordiga}, \citenamefont {Medina}, \citenamefont {Jafarzadeh}, \citenamefont {B{\"o}sch}, \citenamefont {Adams}, \citenamefont {Tournat},\ and\ \citenamefont {Bertoldi}}]{bordigaAutomatedDiscoveryReprogrammable2024}%
  \BibitemOpen
  \bibfield  {author} {\bibinfo {author} {\bibfnamefont {G.}~\bibnamefont {Bordiga}}, \bibinfo {author} {\bibfnamefont {E.}~\bibnamefont {Medina}}, \bibinfo {author} {\bibfnamefont {S.}~\bibnamefont {Jafarzadeh}}, \bibinfo {author} {\bibfnamefont {C.}~\bibnamefont {B{\"o}sch}}, \bibinfo {author} {\bibfnamefont {R.~P.}\ \bibnamefont {Adams}}, \bibinfo {author} {\bibfnamefont {V.}~\bibnamefont {Tournat}},\ and\ \bibinfo {author} {\bibfnamefont {K.}~\bibnamefont {Bertoldi}},\ }\href {https://doi.org/10.1038/s41563-024-02008-6} {\bibfield  {journal} {\bibinfo  {journal} {Nature Materials}\ }\textbf {\bibinfo {volume} {23}},\ \bibinfo {pages} {1486} (\bibinfo {year} {2024})}\BibitemShut {NoStop}%
\bibitem [{\citenamefont {Goel}\ \emph {et~al.}(2024)\citenamefont {Goel}, \citenamefont {Leedumrongwatthanakun}, \citenamefont {Valencia}, \citenamefont {McCutcheon}, \citenamefont {Tavakoli}, \citenamefont {Conti}, \citenamefont {Pinkse},\ and\ \citenamefont {Malik}}]{goelInverseDesignHighdimensional2024}%
  \BibitemOpen
  \bibfield  {author} {\bibinfo {author} {\bibfnamefont {S.}~\bibnamefont {Goel}}, \bibinfo {author} {\bibfnamefont {S.}~\bibnamefont {Leedumrongwatthanakun}}, \bibinfo {author} {\bibfnamefont {N.~H.}\ \bibnamefont {Valencia}}, \bibinfo {author} {\bibfnamefont {W.}~\bibnamefont {McCutcheon}}, \bibinfo {author} {\bibfnamefont {A.}~\bibnamefont {Tavakoli}}, \bibinfo {author} {\bibfnamefont {C.}~\bibnamefont {Conti}}, \bibinfo {author} {\bibfnamefont {P.~W.~H.}\ \bibnamefont {Pinkse}},\ and\ \bibinfo {author} {\bibfnamefont {M.}~\bibnamefont {Malik}},\ }\href {https://doi.org/10.1038/s41567-023-02319-6} {\bibfield  {journal} {\bibinfo  {journal} {Nature Physics}\ }\textbf {\bibinfo {volume} {20}},\ \bibinfo {pages} {232} (\bibinfo {year} {2024})}\BibitemShut {NoStop}%
\bibitem [{\citenamefont {Bennett}\ and\ \citenamefont {Buhmann}(2020)}]{Rob_ID_MQED}%
  \BibitemOpen
  \bibfield  {author} {\bibinfo {author} {\bibfnamefont {R.}~\bibnamefont {Bennett}}\ and\ \bibinfo {author} {\bibfnamefont {S.~Y.}\ \bibnamefont {Buhmann}},\ }\href {https://doi.org/10.1088/1367-2630/abac3a} {\bibfield  {journal} {\bibinfo  {journal} {New Journal of Physics}\ }\textbf {\bibinfo {volume} {22}},\ \bibinfo {pages} {093014} (\bibinfo {year} {2020})}\BibitemShut {NoStop}%
\bibitem [{\citenamefont {Bennett}(2021)}]{bennettInverseDesignEnvironmentinduced2021}%
  \BibitemOpen
  \bibfield  {author} {\bibinfo {author} {\bibfnamefont {R.}~\bibnamefont {Bennett}},\ }\href {https://doi.org/10.1103/PhysRevA.103.013706} {\bibfield  {journal} {\bibinfo  {journal} {Physical Review A}\ }\textbf {\bibinfo {volume} {103}},\ \bibinfo {pages} {013706} (\bibinfo {year} {2021})}\BibitemShut {NoStop}%
\bibitem [{\citenamefont {Matuszak}\ \emph {et~al.}(2022)\citenamefont {Matuszak}, \citenamefont {Buhmann},\ and\ \citenamefont {Bennett}}]{matuszakShapeOptimizationsBodyassisted2022}%
  \BibitemOpen
  \bibfield  {author} {\bibinfo {author} {\bibfnamefont {J.}~\bibnamefont {Matuszak}}, \bibinfo {author} {\bibfnamefont {S.~Y.}\ \bibnamefont {Buhmann}},\ and\ \bibinfo {author} {\bibfnamefont {R.}~\bibnamefont {Bennett}},\ }\href {https://doi.org/10.1103/PhysRevA.106.013515} {\bibfield  {journal} {\bibinfo  {journal} {Physical Review A}\ }\textbf {\bibinfo {volume} {106}},\ \bibinfo {pages} {013515} (\bibinfo {year} {2022})}\BibitemShut {NoStop}%
\bibitem [{\citenamefont {Capers}\ \emph {et~al.}(2021)\citenamefont {Capers}, \citenamefont {Boyes}, \citenamefont {Hibbins},\ and\ \citenamefont {Horsley}}]{capersDesigningCollectiveNonlocal2021}%
  \BibitemOpen
  \bibfield  {author} {\bibinfo {author} {\bibfnamefont {J.~R.}\ \bibnamefont {Capers}}, \bibinfo {author} {\bibfnamefont {S.~J.}\ \bibnamefont {Boyes}}, \bibinfo {author} {\bibfnamefont {A.~P.}\ \bibnamefont {Hibbins}},\ and\ \bibinfo {author} {\bibfnamefont {S.~A.~R.}\ \bibnamefont {Horsley}},\ }\href {https://doi.org/10.1038/s42005-021-00713-1} {\bibfield  {journal} {\bibinfo  {journal} {Communications Physics}\ }\textbf {\bibinfo {volume} {4}},\ \bibinfo {pages} {209} (\bibinfo {year} {2021})}\BibitemShut {NoStop}%
\bibitem [{\citenamefont {{Miguel-Torcal}}\ \emph {et~al.}(2022)\citenamefont {{Miguel-Torcal}}, \citenamefont {{Abad-Arredondo}}, \citenamefont {{Garcia-Vidal}},\ and\ \citenamefont {{Fernandez-Dominguez}}}]{miguel-torcalInversedesignedDielectricCloaks2022a}%
  \BibitemOpen
  \bibfield  {author} {\bibinfo {author} {\bibfnamefont {A.}~\bibnamefont {{Miguel-Torcal}}}, \bibinfo {author} {\bibfnamefont {J.}~\bibnamefont {{Abad-Arredondo}}}, \bibinfo {author} {\bibfnamefont {F.~J.}\ \bibnamefont {{Garcia-Vidal}}},\ and\ \bibinfo {author} {\bibfnamefont {A.~I.}\ \bibnamefont {{Fernandez-Dominguez}}},\ }\href {https://doi.org/10.1515/nanoph-2022-0231} {\bibfield  {journal} {\bibinfo  {journal} {Nanophotonics}\ }\textbf {\bibinfo {volume} {11}},\ \bibinfo {pages} {4387} (\bibinfo {year} {2022})}\BibitemShut {NoStop}%
\bibitem [{\citenamefont {Cisowski}\ \emph {et~al.}(2024)\citenamefont {Cisowski}, \citenamefont {Waller},\ and\ \citenamefont {Bennett}}]{ID_Rob_Claire}%
  \BibitemOpen
  \bibfield  {author} {\bibinfo {author} {\bibfnamefont {C.~M.}\ \bibnamefont {Cisowski}}, \bibinfo {author} {\bibfnamefont {M.~C.}\ \bibnamefont {Waller}},\ and\ \bibinfo {author} {\bibfnamefont {R.}~\bibnamefont {Bennett}},\ }\href {https://doi.org/10.1103/PhysRevA.109.043533} {\bibfield  {journal} {\bibinfo  {journal} {Phys. Rev. A}\ }\textbf {\bibinfo {volume} {109}},\ \bibinfo {pages} {043533} (\bibinfo {year} {2024})}\BibitemShut {NoStop}%
\bibitem [{\citenamefont {Miller}(2012)}]{MillerPhD2012}%
  \BibitemOpen
  \bibfield  {author} {\bibinfo {author} {\bibfnamefont {O.~D.}\ \bibnamefont {Miller}},\ }\href@noop {} {\bibfield  {journal} {\bibinfo  {journal} {Ph. D. Thesis}\ } (\bibinfo {year} {2012})}\BibitemShut {NoStop}%
\bibitem [{\citenamefont {Buhmann}\ and\ \citenamefont {Welsch}(2006)}]{buhmannBornExpansionCasimirPolder2006}%
  \BibitemOpen
  \bibfield  {author} {\bibinfo {author} {\bibfnamefont {S.~Y.}\ \bibnamefont {Buhmann}}\ and\ \bibinfo {author} {\bibfnamefont {D.-G.}\ \bibnamefont {Welsch}},\ }\href {https://doi.org/10.1007/s00340-005-2055-3} {\bibfield  {journal} {\bibinfo  {journal} {Applied Physics B}\ }\textbf {\bibinfo {volume} {82}},\ \bibinfo {pages} {189} (\bibinfo {year} {2006})}\BibitemShut {NoStop}%
\bibitem [{\citenamefont {Rodriguez}\ \emph {et~al.}(2007)\citenamefont {Rodriguez}, \citenamefont {Ibanescu}, \citenamefont {Iannuzzi}, \citenamefont {Joannopoulos},\ and\ \citenamefont {Johnson}}]{Rodrig2007}%
  \BibitemOpen
  \bibfield  {author} {\bibinfo {author} {\bibfnamefont {A.}~\bibnamefont {Rodriguez}}, \bibinfo {author} {\bibfnamefont {M.}~\bibnamefont {Ibanescu}}, \bibinfo {author} {\bibfnamefont {D.}~\bibnamefont {Iannuzzi}}, \bibinfo {author} {\bibfnamefont {J.~D.}\ \bibnamefont {Joannopoulos}},\ and\ \bibinfo {author} {\bibfnamefont {S.~G.}\ \bibnamefont {Johnson}},\ }\href {https://doi.org/10.1103/PhysRevA.76.032106} {\bibfield  {journal} {\bibinfo  {journal} {Phys. Rev. A}\ }\textbf {\bibinfo {volume} {76}},\ \bibinfo {pages} {032106} (\bibinfo {year} {2007})}\BibitemShut {NoStop}%
\bibitem [{\citenamefont {Rodriguez}\ \emph {et~al.}(2009)\citenamefont {Rodriguez}, \citenamefont {McCauley}, \citenamefont {Joannopoulos},\ and\ \citenamefont {Johnson}}]{Rodrig2009}%
  \BibitemOpen
  \bibfield  {author} {\bibinfo {author} {\bibfnamefont {A.~W.}\ \bibnamefont {Rodriguez}}, \bibinfo {author} {\bibfnamefont {A.~P.}\ \bibnamefont {McCauley}}, \bibinfo {author} {\bibfnamefont {J.~D.}\ \bibnamefont {Joannopoulos}},\ and\ \bibinfo {author} {\bibfnamefont {S.~G.}\ \bibnamefont {Johnson}},\ }\href {https://doi.org/10.1103/PhysRevA.80.012115} {\bibfield  {journal} {\bibinfo  {journal} {Phys. Rev. A}\ }\textbf {\bibinfo {volume} {80}},\ \bibinfo {pages} {012115} (\bibinfo {year} {2009})}\BibitemShut {NoStop}%
\bibitem [{\citenamefont {Kristensen}\ \emph {et~al.}(2023)\citenamefont {Kristensen}, \citenamefont {Beverungen}, \citenamefont {Intravaia},\ and\ \citenamefont {Busch}}]{Francesco}%
  \BibitemOpen
  \bibfield  {author} {\bibinfo {author} {\bibfnamefont {P.~T.}\ \bibnamefont {Kristensen}}, \bibinfo {author} {\bibfnamefont {B.}~\bibnamefont {Beverungen}}, \bibinfo {author} {\bibfnamefont {F.}~\bibnamefont {Intravaia}},\ and\ \bibinfo {author} {\bibfnamefont {K.}~\bibnamefont {Busch}},\ }\href {https://doi.org/10.1103/PhysRevB.108.205424} {\bibfield  {journal} {\bibinfo  {journal} {Phys. Rev. B}\ }\textbf {\bibinfo {volume} {108}},\ \bibinfo {pages} {205424} (\bibinfo {year} {2023})}\BibitemShut {NoStop}%
\bibitem [{\citenamefont {Oskooi}\ \emph {et~al.}(2010)\citenamefont {Oskooi}, \citenamefont {Roundy}, \citenamefont {Ibanescu}, \citenamefont {Bermel}, \citenamefont {Joannopoulos},\ and\ \citenamefont {Johnson}}]{Meep}%
  \BibitemOpen
  \bibfield  {author} {\bibinfo {author} {\bibfnamefont {A.~F.}\ \bibnamefont {Oskooi}}, \bibinfo {author} {\bibfnamefont {D.}~\bibnamefont {Roundy}}, \bibinfo {author} {\bibfnamefont {M.}~\bibnamefont {Ibanescu}}, \bibinfo {author} {\bibfnamefont {P.}~\bibnamefont {Bermel}}, \bibinfo {author} {\bibfnamefont {J.}~\bibnamefont {Joannopoulos}},\ and\ \bibinfo {author} {\bibfnamefont {S.~G.}\ \bibnamefont {Johnson}},\ }\href {https://doi.org/https://doi.org/10.1016/j.cpc.2009.11.008} {\bibfield  {journal} {\bibinfo  {journal} {Computer Physics Communications}\ }\textbf {\bibinfo {volume} {181}},\ \bibinfo {pages} {687} (\bibinfo {year} {2010})}\BibitemShut {NoStop}%
\bibitem [{Note2()}]{Note2}%
  \BibitemOpen
  \bibinfo {note} {The term forward is conventionally defined as opposite-to-adjoint, and \protect \textit {not} as forward-time; however, this interpretation applies in this work.}\BibitemShut {Stop}%
\bibitem [{\citenamefont {Osher}\ and\ \citenamefont {Sethian}(1988)}]{LevelSetOSHER}%
  \BibitemOpen
  \bibfield  {author} {\bibinfo {author} {\bibfnamefont {S.}~\bibnamefont {Osher}}\ and\ \bibinfo {author} {\bibfnamefont {J.~A.}\ \bibnamefont {Sethian}},\ }\href {https://doi.org/https://doi.org/10.1016/0021-9991(88)90002-2} {\bibfield  {journal} {\bibinfo  {journal} {Journal of Computational Physics}\ }\textbf {\bibinfo {volume} {79}},\ \bibinfo {pages} {12} (\bibinfo {year} {1988})}\BibitemShut {NoStop}%
\bibitem [{\citenamefont {Guyer}\ \emph {et~al.}(2009)\citenamefont {Guyer}, \citenamefont {Wheeler},\ and\ \citenamefont {Warren}}]{Fipy}%
  \BibitemOpen
  \bibfield  {author} {\bibinfo {author} {\bibfnamefont {J.~E.}\ \bibnamefont {Guyer}}, \bibinfo {author} {\bibfnamefont {D.}~\bibnamefont {Wheeler}},\ and\ \bibinfo {author} {\bibfnamefont {J.~A.}\ \bibnamefont {Warren}},\ }\href {https://doi.org/10.1109/MCSE.2009.52} {\bibfield  {journal} {\bibinfo  {journal} {Computing in Science \& Engineering}\ }\textbf {\bibinfo {volume} {11}},\ \bibinfo {pages} {6} (\bibinfo {year} {2009})}\BibitemShut {NoStop}%
\bibitem [{\citenamefont {Stein}\ and\ \citenamefont {contributors}(2024)}]{scikit-fmm}%
  \BibitemOpen
  \bibfield  {author} {\bibinfo {author} {\bibfnamefont {D.}~\bibnamefont {Stein}}\ and\ \bibinfo {author} {\bibnamefont {contributors}},\ }\href {https://github.com/scikit-fmm/scikit-fmm} {\bibinfo {title} {scikit-fmm: The fast marching method for python}} (\bibinfo {year} {2024}),\ \bibinfo {note} {accessed: 2024-07-01}\BibitemShut {NoStop}%
\end{thebibliography}%

\end{document}